\documentclass[pra,twocolumn,showpacs,amsmath,amsfonts]{revtex4}
\usepackage{bm}
\usepackage{graphicx}

\newcommand{\bra}[1]{\langle #1 | \,}
\newcommand{\ket}[1]{\, | #1 \rangle}

\newcommand{\lra}{\leftrightarrow}
\newcommand{\la}{\lambda}
\newcommand{\om}{\omega}
\newcommand{\Om}{\Omega}
\newcommand{\ga}{\gamma}
\newcommand{\Ga}{\Gamma}
\newcommand{\de}{\delta}
\newcommand{\De}{\Delta}

\begin{document}

\title{Quantum computer with dipole-dipole interacting two-level systems}

\author{David Petrosyan}
\email[E-mail: ]{dap@iesl.forth.gr}
\affiliation{Institute of Electronic Structure \& Laser, 
FORTH, Heraklion 71110, Crete, Greece}
\author{Gershon Kurizki}
\email[E-mail: ]{gershon.kurizki@weizmann.ac.il}
\affiliation{Department of Chemical Physics, 
Weizmann Institute of Science, Rehovot 76100, Israel}

\date{\today}

\begin{abstract}
A scalable, high-performance quantum processor can be implemented using
near-resonant dipole-dipole interacting dopants in a solid state host. 
In this scheme, the qubits are represented by ground and subradiant states 
of effective dimers formed by pairs of closely spaced two-level systems, 
while the two-qubit entanglement either relies on the coherent excitation 
exchange between the dimers or is mediated by external laser fields.
\end{abstract}


\maketitle

\section{Introduction}

Quantum information science, which is based on quantum principles 
and which extends and generalizes the classical information theory, is 
currently attracting enormous interest, due to its fundamental nature 
and its potentially revolutionary applications to computation and 
secure communication \cite{QCQI}. Quantum information processing schemes
rely on the ability to coherently manipulate and couple (or entangle) 
the qubits---quantum analogs of classical bits. The main stumbling blocks 
en route to the realization of useful quantum computers, comprised of many 
qubits, are: (i) {\em fidelity loss} due to decoherence, 
which grows with the amount of single- and two-qubit operations and 
requires large redundancy for the application of error-correction methods 
\cite{errCorr,ECnnch}; (ii) {\em scalability} of the quantum processor, 
which restricts the choice of candidate systems. Although the various 
proposals and experimental demonstrations of rudimentary quantum computers 
have thus far predominantly involved optical manipulations of atoms in ion 
traps \cite{iontr}, high-Q cavities \cite{PGCZ,MBBKNSK}, optical lattices 
\cite{BCJD} and microtraps \cite{JCZ,CZnat,SDKMRM}, the low fidelity and/or 
difficulties with scalability of these schemes cast doubts on their 
suitability for truly large-scale quantum computation. Solid-state quantum
processors with quantum dots \cite{qdots,IABDVLShS,BLW} or active dopants 
\cite{dopsSC,BDHT,LH} appear to be more promising, both principally 
and technologically.

In a recent publication \cite{PKprl} we have proposed a combined 
optical/solid-state approach that can significantly improve the speed, 
fidelity and scalability of a quantum processor. The crux of this approach 
is the concept of dimer qubit, wherein two similar two-level systems 
(e.g., quantum dots), separated by a few nanometers and interacting with
each other via the resonant dipole-dipole interaction (RDDI) 
\cite{Lehmberg,Agarwal-MQE,GKABR}, form an effective dimer, whose ground 
and subradiant \cite{Dicke} states serve as robust qubit states. It is the 
purpose of this paper to give a complete account of that scheme and 
compare it with other related systems proposed for physical realization of 
quantum computation. We will show that all the basic ingredients of 
quantum computation \cite{DiVincenzo}, i.e., state preparation, universal 
logic gates and qubit readout, are realizable by optical manipulations 
of these dimers. A scalable quantum processor is envisioned in a 
cryogenically-cooled solid-state host material doped with such dimers 
at controllable nanoscale separations.

The paper is organized as follows. In Sec. \ref{sec:qubit} we review the
resonant dipole-dipole interaction between two two-level systems (atoms),
which build our dimer qubit. We then discuss the laser-dimer interaction
and outline the mechanism by which the qubits are manipulated and measured. 
In Sec. \ref{sec:ent} we study the dipole-dipole interaction between pairs
of closely spaced dimers, which mediates their entanglement. Finally,
in Sec. \ref{sec:impl} we describe the implementation of the scalable
quantum processor, followed by the concluding remarks.

\section{The Qubit}
\label{sec:qubit}

In this section we will introduce the notion of ``dimer qubit'' and outline 
the principles of its manipulation and measurement. 

\subsection{Resonant dipole-dipole interaction}

\begin{figure}[t]
\centerline{\includegraphics[width=8.5cm]{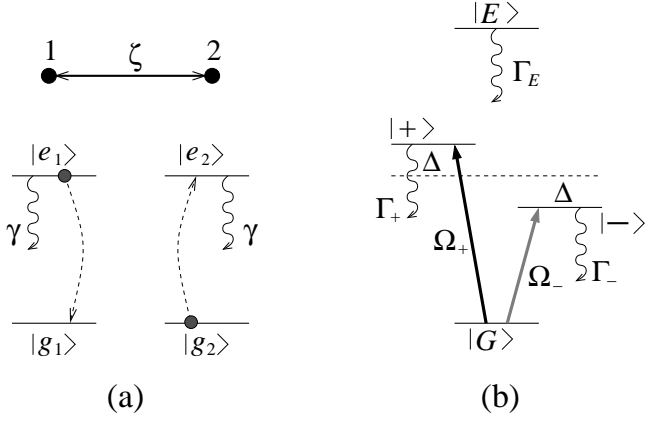}}
\caption{(a) Two TLAs 1 and 2, separated by normalized distance $\zeta$,
interact via RDDI and exchange a single excitation.
(b) Energy level diagram of the resulting ``dimer'' states of the system.
\label{fig:dd_int}}
\end{figure}

Let us first recall the cooperative properties of two identical two-level 
atoms (TLAs), 1 and 2, at fixed positions $\textbf{r}_1$ and $\textbf{r}_2$, 
whose ground and excited states are labeled as $\ket{g_{1,2}}$ and 
$\ket{e_{1,2}}$, respectively [Fig.~\ref{fig:dd_int}(a)]. The atoms interact 
with each other via the continuum of free-space modes of electromagnetic field.
Using the standard Born-Markov approximation to eliminate the vacuum modes of 
the photonic continuum \cite{Lehmberg,Agarwal-MQE}, one can derive 
an effective non-Hermitian Hamiltonian for the system of two TLAs, which
can be cast in a form 
\begin{equation}
H = H_{\text{Atom}}+V_{\text{RDDI}} \label{ham}.
\end{equation}
Here the first term 
\begin{equation}
H_{\text{Atom}} = \hbar (\om_{eg} - i \ga/2) 
(\ket{e_1} \bra{e_1} + \ket{e_2} \bra{e_2})
\end{equation} 
represents the atomic Hamiltonian, where $\om_{eg}$ is the resonant frequency 
and $\ga$ the radiative decay rate on the atomic transition 
$\ket{e}\to \ket{g}$. The second term 
\begin{equation}
V_{\text{RDDI}} = \hbar \left(\De - i \ga_{12}/2 \right)
(\ket{e_1g_2} \bra{g_1e_2} + \ket{g_1 e_2} \bra{e_1 g_2}) , \label{rddi}
\end{equation}
with
\begin{eqnarray}
\De &=& \frac{3\ga}{4}\left\{-[1-\cos^2\theta ] 
\frac{\cos{\zeta}}{\zeta} 
\right. \nonumber \\ & & \;\;\;\;\;\;\;\; \left.
+[1-3 \cos^2\theta ] 
\left( \frac{\sin{\zeta}}{\zeta^2} + 
\frac{\cos{\zeta}}{\zeta^3}
\right)\right\} , \\
\ga_{12} &=&  \frac{3\ga}{2}\left\{[1-\cos^2\theta  ] 
\frac{\sin{\zeta}}{\zeta} 
\right. \nonumber \\ & & \;\;\;\;\;\;\;\; \left.
+[1-3 \cos^2\theta  ] 
\left( \frac{\cos{\zeta}}{\zeta^2} - 
\frac{\sin{\zeta}}{\zeta^3}
\right)\right\} , 
\end{eqnarray} 
describes the resonant dipole-dipole interaction (RDDI) between the atoms,
where $\theta$ is the angle between the direction of atomic dipole moment
and the interatomic axis, and $\zeta = q r_{12}$, with $q = \om_{eg}/c$ 
and $r_{12} = |\textbf{r}_1 - \textbf{r}_2|$, is the normalized distance
between the atoms. Thus, the rate of coherent excitation exchange 
between the atoms is given by the real part of the RDDI potential $\De$, 
while the imaginary part of the potential $\ga_{12}$ is responsible
for the cooperative radiative decay of the system. When $\zeta \gg 1$,
both $\De$ and $\ga_{12}$ vanish and so does the $V_{\text{RDDI}}$, and 
we essentially deal with the system of two independent atoms described by 
the Hamiltonian $H = H_{\text{Atom}}$.

In the opposite limit of small interatomic separations $\zeta < 1$, 
the natural basis of sates for the two-atom system is the molecular basis. 
The transformation form the atomic to molecular basis is achieved via 
the diagonalization of the Hamiltonian (\ref{ham}). This yields the 
dressed by the RDDI ``dimer'' eigenstates 
\begin{equation}
\begin{array}{c}
\ket{G} = \ket{g_1 g_2}, \;\; \ket{E} = \ket{e_1 e_2} , \\
\\
\ket{\pm} = \frac{1}{\sqrt{2}}(\ket{e_1 g_2} \pm \ket{g_1 e_2}), 
\end{array} \label{deigst}
\end{equation}
with the corresponding eigenvalues
\[
\begin{array}{c}
\la_G = 0 , \;\; \la_E = 2 \om_{eg} -i \Ga_E/2 , \\
\\
\la_{\pm} = \om_{eg} \pm \De -i \Ga_{\pm}/2 ,
\end{array}
\] 
as shown in Fig.~\ref{fig:dd_int}(b). 
Thus the symmetric (superradiant) $\ket{+}$ and doubly-excited 
$\ket{E}$ eigenstates have corresponding decay rates 
$\Ga_+ = \ga+\ga_{12}$ and $\Ga_E = 2 \ga$ which exceed that of a single
isolated atom, while the decay rate $\Ga_- = \ga - \ga_{12}$ of the 
antisymmetric (subradiant) eigenstate $\ket{-}$ is suppressed \cite{Dicke}.
For very small interatomic separations $\zeta \ll 1$ and $\theta = \pi/2$, 
the real part of the RDDI potential $\De$ and decay rates of the 
corresponding dimer states can be approximated as 
$\De \approx 3 \ga/(4 \zeta^3) \gg \ga$, $\Ga_+ \approx  \Ga_E = 2 \ga$, 
and $\Ga_- \approx \ga \zeta^2/5 \ll \ga$.

\subsection{Dimer-laser interaction}
\label{dlint}

Let us irradiate the pair of atoms with a laser field 
$E(\textbf{r},t) = \mathcal{E} e^{i (\textbf{k} \textbf{r} -\om t)}$ 
having frequency $\om \sim \om_{eg}$, wave vector $\textbf{k}$, and 
phase $\varphi$ ($\mathcal{E} = |\mathcal{E}| e^{i \varphi} $). 
The Hamiltonian (\ref{ham}) for the system of two atoms acquires
now an atom-field interaction term
\begin{equation}
V_{\text{AFI}} = \hbar \Om e^{-i \om t} 
(e^{i \textbf{k} \textbf{r}_1} \ket{e_1}\bra{g_1} 
+ e^{i \textbf{k} \textbf{r}_2} \ket{e_2}\bra{g_2}) 
+ \text{H. c.} \label{afint} ,
\end{equation}
where $\Om = \mu \mathcal{E}/\hbar$ is the Rabi frequency of the field for 
a single isolated atom, $\mu$ being the dipole matrix element for the atomic 
transition $\ket{g} \to \ket{e}$. Then, in the RDDI-dressed basis 
(\ref{deigst}), the interaction-picture Hamiltonian takes the form
\begin{widetext}
\begin{eqnarray}
H_{\rm int}/\hbar &=& (\de - \De - i \Ga_- /2) \ket{-} \bra{-} 
+ (\de + \De - i \Ga_+ /2) \ket{+} \bra{+} 
+ (2 \de - i \Ga_E /2) \ket{E} \bra{E} 
\nonumber \\ & &
+ \Om_{-} (\ket{-}\bra{G} - \ket{E}\bra{-}) 
+ \Om_{+} (\ket{+}\bra{G} + \ket{E}\bra{+}) + \text{H. c.} , \label{DIHam} 
\end{eqnarray}
\end{widetext}
where $\de = \om_{eg} - \om$ is the laser field detuning 
from the $\ket{g} \to \ket{e}$ transition resonance and 
\[
\Om_{\pm} = \frac{1}{\sqrt{2}} \, 
\Om \, [1 \pm e^{-i \textbf{k} \textbf{r}_{12}}]. 
\]
Thus, the Rabi frequencies (coupling strengths) of the laser field on the 
dimer transitions $\ket{G} \to \ket{-}$ and $\ket{-} \to \ket{E}$ are equal 
to $\pm \Om_{-}$, respectively, and on the transitions $\ket{G} \to \ket{+}$ 
and $\ket{+} \to \ket{E}$ are equal to $\Om_{+}$. In the limit of small 
interatomic separations, $r_{12} \ll k^{-1}$, one has 
\[ 
\Om_+ \simeq \sqrt{2} \Om , \;\;\; 
\Om_- \simeq i \frac{\Om}{\sqrt{2}} \, \zeta \cos \phi,
\] 
where $\phi$ is the angle between the vectors $\textbf{k}$ and 
$\textbf{r}_{12}$. Note that $\Om_-$ identically vanishes if the laser 
field propagation direction is perpendicular to the interatomic axis, 
$\textbf{k} \perp \textbf{r}_{12}$, while, in the case of $\zeta \ll 1$, 
it is maximized for the $\textbf{k} \parallel \textbf{r}_{12}$ configuration.
In physical terms, the subradiant $\ket{G} \to \ket{-}$ transition exhibits 
a quadrupolar behavior and dipole-moment suppression, due to destructive 
interference of the two-atom interactions with the field, as opposed to 
their constructive interference in the superradiant $\ket{G} \to \ket{+}$ 
transition. 

Consider first the case of $\de \simeq \De$, i.e., the frequency of 
electric field $\om$ is resonant with the dimer transition 
$\ket{G} \to \ket{-}$ [Fig. \ref{fig:dd_int}(b)]. If the initial state of the
dimer is either $\ket{G}$ or $\ket{-}$, the transitions $\ket{G} \to \ket{+}$
and $\ket{-} \to \ket{E}$ are nonresonant, as they are detuned by $2 \De$. 
Then, provided $\De \gg \ga, \Om$, we can adiabatically eliminate the 
nonresonant states $\ket{+}$ and $\ket{E}$, obtaining an effective two-level
system described by the Hamiltonian
\begin{eqnarray}
H_{\rm eff}^{(-)}/\hbar &=&  - i \Ga_G /2 \ket{G} \bra{G} - 
i \Ga_- /2 \ket{-} \bra{-} \nonumber \\ & &
+ \Om_{-} \ket{-}\bra{G} + \text{H. c.} \label{effQHam} 
\end{eqnarray}
Here $\Ga_G = \Ga_+ |\Om_+|^2/(2\De)^2$ is the relaxation rate of the ground
state $\ket{G}$ due to its nonresonant coupling with the superradiant state
$\ket{+}$, while the residual Stark shifts $|\Om_{+}|^2/(2\De)$ and 
$|\Om_{-}|^2/(2\De)$ of levels $\ket{G}$ and $\ket{-}$, respectively, are 
absorbed in the laser field detuning, 
$\de = \De -(|\Om_{+}|^2 -|\Om_{-}|^2)/(2\De) \simeq \De$. Assuming that 
$|\Om_-| \gg \Ga_-$ and $|\Om_+| \ll 2 \De$, i.e., 
$\zeta \ll |\Om|/\ga \ll \zeta^{-3}$, and therefore 
neglecting for the moment the relaxation terms in (\ref{effQHam}), the 
resulting evolution operator 
$U(T) = \exp \left(-\frac{i}{\hbar} \int_0^{T} H_{\rm eff}^{(-)} d t\right)$ 
takes the familiar form
\[
U(T) =\left[ \begin{array}{cc}
\cos |\Om_-| T & -i e^{i \varphi} \sin  |\Om_-| T  \\ 
-i e^{- i \varphi}\sin |\Om_-| T &  \cos  |\Om_-| T  
\end{array} \right] ,
\]
which describes the coherent Rabi osculations between levels $\ket{G}$ and 
$\ket{-}$ with frequency $\Om_{-}$.

Consider now the opposite case $\de \simeq - \De$, when the frequency 
of electric field $\om$ is resonant with the dimer transition 
$\ket{G} \to \ket{+}$ [Fig. \ref{fig:dd_int}(b)]. For the dimer initially 
in state $\ket{G}$, the transition $\ket{G} \to \ket{-}$ is detuned by 
$2 \De \gg \Ga_-, \Om_- $. Similarly to the previous case, we can then 
neglect the nonresonant transitions $\ket{G} \to \ket{-}$ and 
$\ket{+} \to \ket{E}$, obtaining the effective Hamiltonian
\begin{equation}
H_{\rm eff}^{(+)}/\hbar =  - i \Ga_+ /2 \ket{+} \bra{+} 
+ \Om_{+} \ket{+}\bra{G} + \text{H. c.} \label{effMHam} 
\end{equation}
Since the decay rate of the superradiant state is large, 
$\Ga_+ \approx 2 \ga$, for moderate field amplitudes $\Om \sim \ga$
the coherent Rabi oscillations on the transition $\ket{G} \to \ket{+}$
will not persist. Rather, the system will very quickly settle to the 
steady-state, in which the populations of states $\ket{+}$ and $\ket{G}$ 
are given, respectively, by
\[
\rho_{++} \simeq \frac{|\Om_{+}|^2} {(\Ga_{+}/2)^2 + 2 |\Om_{+}|^2} , \;\;\;
\rho_{GG} \simeq 1 - \rho_{++}.
\]

\subsection{Single-qubit rotations}
\label{sec:sqr}

\begin{figure}[t]
\centerline{\includegraphics[width=8.5cm]{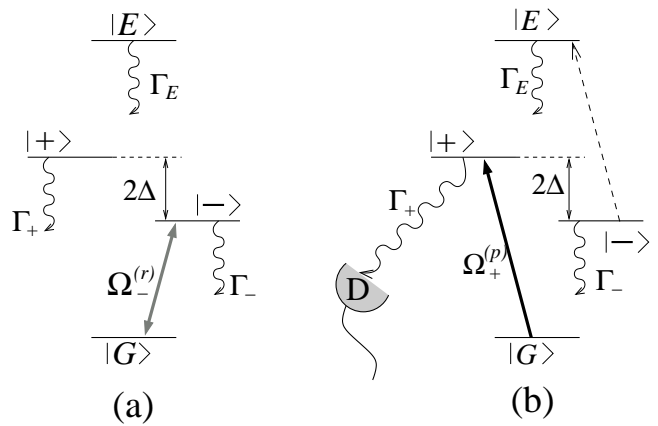}}
\caption{(a) Single-qubit rotations are realized by the laser
field $\Om_-^{(r)}$ resonant with the qubit transition 
$\ket{G} \lra \ket{-}$ of the dimer.
(b) During the readout, if the qubit state is $\ket{G}$, the laser field 
$\Om_+^{(p)}$ resonant with the dimer transition $\ket{G} \to \ket{+}$ 
induces strong fluorescence, which is monitored by the detector D.
\label{fig:rotread}}
\end{figure}

Now we are in a position to introduce the concept of subradiant dimer 
qubit. The two basis states of the qubit are represented by the dimer 
ground $\ket{G}$ and subradiant $\ket{-}$ states. An arbitrary single-qubit 
operation (rotation) can be performed by the laser field 
$E_r = \mathcal{E}_r e^{i (\textbf{k}_r \textbf{r} -\om_r t)}$
with wave vector $\textbf{k}_r \parallel \textbf{r}_{12}$ and frequency 
$\om_r = \om_{eg} -\De$ that is resonant with the qubit transition 
$\ket{G} \to \ket{-}$ [Fig.~\ref{fig:rotread}(a)]. As an example, a laser 
pulse of area $\Om_-^{(r)} T = \pi/2$ ($\pi$-pulse \cite{c_pi-pulse}) would 
realize the {\sc not} gate that interchanges the qubit states 
$\ket{G} \lra \ket{-}$. Similarly, a pulse with the phase $\varphi = -\pi /2$
and area $\Om_-^{(r)} T = \pi/4$ ($\pi/2$-pulse) would realize, to within the
phase-flip of state $\ket{-}$, the Hadamard transformation. 

It is easy to estimate the error per single-qubit rotation operation. 
During the qubit flip time $T_{\text{flip}} = \pi / (2 |\Om_-^{(r)}|)$, 
the error probability due to the spontaneous emission from the subradiant 
state $\ket{-}$ has the upper bound  
$P_-^{\text{sp}} \leq \Ga_- T_{\text{flip}}=\pi \ga \zeta/(5 \sqrt{2} \Om_r)$,
while the error probability due to the population transfer from the ground 
state $\ket{G}$ to the superradiant state $\ket{+}$ satisfies 
$P_+^{\text{tr}} \leq \Ga_+ |\Om_+^{(r)}|^2 T_{\text{flip}} /(2 \De)^2 = 
8\sqrt{2} \pi \Om_r \zeta^5/(9 \ga)$. Minimizing the total error probability
$P_{\text{qubit}} = P_-^{\text{sp}} + P_+^{\text{tr}}$ with respect to
$\Om_r$, we find that, given the values of $\ga$ and $\zeta$, the smallest 
error per gate operation is attained for $\Om_r/\ga \simeq (3 \zeta^2)^{-1}$, 
for which $P_{\text{qubit}}^{\text{min}} \leq 2\ga/\De \simeq 2.65 \zeta^3$.
As an example, for the parameters $\zeta \simeq 0.033$ and 
$\Om_r/\ga \simeq 300$, the RDDI strength is $\De \approx 2\times 10^{4} \ga$, 
the decay rate of the antisymmetric state is 
$\Ga_- \approx 2 \times 10^{-4} \ga$, and the error probability during the 
qubit flip-time is $P_{\text{qubit}}^{\text{min}} \leq 10^{-4}$, as compared 
to the corresponding error probability for a single atom 
$P_{\text{atom}}^{\text{sp}} \leq \pi \ga /(2|\Om_r|) \simeq 5 \times 10^{-3}$.
Such small memory and gate operation errors are amenable to error correction 
\cite{errCorr,ECnnch}. 

\begin{figure}[t]
\centerline{\includegraphics[width=3.8cm]{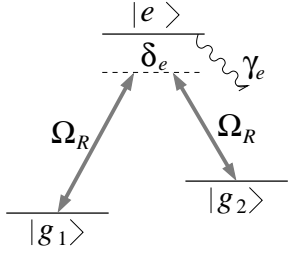}}
\caption{Schematic representation of an atomic system where the qubit
basis states correspond to the long-lived ground states $\ket{g_1}$ and 
$\ket{g_2}$, and single qubit rotations are performed via two Raman
fields acting on the transitions $\ket{g_1} \to \ket{e}$ and
$\ket{e} \to \ket{g_2}$. 
\label{fig:R_tr}}
\end{figure}

Let us compare the present scheme with another common scheme 
\cite{BCJD,LH,PGCZ,IABDVLShS}, where the qubit basis states are represented 
by two metastable ground states $\ket{g_1}$ and $\ket{g_2}$ of an isolated 
atom (Fig. \ref{fig:R_tr}). The single-qubit rotations are performed by two 
laser fields tuned to the two-photon Raman resonance between the two ground
states. In order to minimize the losses, these fields are strongly detuned 
from the excited state $\ket{e}$ by detuning $\de_e \gg \ga_e$, where $\ga_e$ 
is the spontaneous decay rate of that state. For simplicity, let us assume 
that the Rabi frequencies of both Raman fields are equal to $\Om_R$. Then the 
effective Rabi frequency on the two-photon transition $\ket{g_1} \to \ket{g_2}$
is given by $\Om_{\text{eff}}^{(R)} = \Om_R^2/\de_e$. Hence, during the qubit 
flip time $T_{\text{flip}}^{(R)} = \pi / (2 \Om_{\text{eff}}^{(R)})$, for the 
error probability due to the decay from the excited state $\ket{e}$ 
we obtain $P_e^{\text{sp}} \leq \ga_e |\Om_R|^2 T_{\text{flip}}^{(R)}/\de_e^2= 
\pi \ga_e /(2\de_e)$. With the parameters similar to those for the dimer qubit,
$\Om_{\text{eff}}^{(R)} \simeq \Om_-^{(r)}$ and $\Om_R/\ga_e \simeq 300$, 
which yields $\de_e \approx 1.3 \times 10^4 \ga_e$, we obtain that 
$P_e^{\text{sp}} \simeq 1.2 \times 10^{-4}$, which is comparable to the
error probability in our scheme. However, the dimer qubit based on the RDDI 
between TLAs is easier to manipulate since it requires only one laser field.

\subsection{Qubit measurement}

Next we outline the readout scheme. The method we propose is similar to the 
electron shelving or quantum jump technique \cite{QJumps}. In order 
to measure the state of the qubit, we irradiate it with the probe field 
$E_p = \mathcal{E}_p e^{i (\textbf{k}_p \textbf{r} -\om_p t)}$ having 
frequency $\om_p = \om_{eg} + \De$ that is resonant with the dimer 
transition $\ket{G} \to \ket{+}$, and collect the fluorescence signal from 
the superradiant state $\ket{+}$ [Fig.~\ref{fig:rotread}(b)]. The Rabi 
frequency on that transition is $\Om_+^{(p)}$, while on the qubit transition 
$\ket{G} \to \ket{-}$, from which the probe field is detuned by 
$2\De \gg \Om_-^{(p)}$, its Rabi frequency $\Om_-^{(p)}$ is much smaller. 
Therefore the presence or absence of fluorescence will indicate whether the 
qubit is in the ``bright'' state $\ket{G}$ or in the ``dark'' state $\ket{-}$. 
With the qubit in state $\ket{G}$, the probability of detecting 
the fluorescence by a detector with finite efficiency $\eta < 1$ during the 
time $T_{\text{pr}}$ the probe field is on, is given by 
\begin{equation}
P_G^{\text{fl}} = \eta \Ga_+ \frac{|\Om_{+}^{(p)}|^2}
{(\Ga_{+}/2)^2 + 2 |\Om_{+}^{(p)}|^2} T_{\text{pr}} , 
\end{equation}
which, in the case of $\Om_p > \ga$, can be approximated as 
$P_G^{\text{fl}} \approx \eta \ga T_{\text{pr}}$. Requiring that 
$P_G^{\text{fl}} \simeq 1$, we obtain for the detection time 
$T_{\text{pr}} \simeq (\eta \ga)^{-1}$, which, on the other hand, 
should be much smaller than the lifetime of the qubit state $\ket{-}$, 
$T_{\text{pr}} \ll \Ga_-^{-1}$. 
 
It is imperative to note, however, that the probe frequency $\om_p$ 
exactly matches that of the transition $\ket{-} \to \ket{E}$, on which the 
probe field Rabi frequency is $-\Om_-^{(p)}$ [see Eq.~(\ref{DIHam})]. 
Therefore, the dimer in state $\ket{-}$ can first be excited to $\ket{E}$ by 
absorbing a probe photon, then decay to $\ket{+}$, subsequently producing 
the same fluorescence signal as if it were initially in state $\ket{G}$. 
One can show that the probability of detecting the fluorescence in that 
case is described by the equation 
\begin{equation}
P_-^{\text{fl}} = \eta \Ga_+ \frac{|\Om_{+}^{(p)}|^2}
{(\Ga_{+}/2)^2 + 2 |\Om_+^{(p)}|^2} \int_0^{T_{\text{pr}}} 
[1 - e^{-\ga_{-+} t}] dt , 
\end{equation}
which, with $\ga_{-+} T_{\text{pr}} < 1$, can be approximated as 
$P_-^{\text{fl}} \approx \eta \ga \ga_{-+} T_{\text{pr}}^2/2$, where
$\ga_{-+} = 4 |\Om_-^{(p)}|^2 /\Ga_E = |\Om_p|^2 \zeta^2/\ga$ is the rate of 
transition $\ket{-} \to \ket{+}$ due to the absorption of a probe photon and 
consequent decay from the state $\ket{E}$. Requiring that 
$P_-^{\text{fl}} \ll 1$, while still $P_G^{\text{fl}} \simeq 1$, we obtain 
the following condition on the system parameters, 
$\Om_p /\ga < \sqrt{2 \eta} /\zeta$. With a realistic $\eta \simeq 0.3 $ and 
$\Om_p /\ga \simeq 3$ ($\zeta \simeq 0.033$), for the reliability of the 
measurement we obtain 
\[
\frac{P_G^{\text{fl}}}{P_G^{\text{fl}} + P_-^{\text{fl}}} = 
\frac{2 \eta}{2 \eta + (\Om_p \zeta /\ga)^2} = 98\% .
\]
We finally note that if the propagation direction of the probe field is 
perpendicular to the interatomic axis, then $\Om_-^{(p)} = 0$ and the above 
idle fluorescence does not occur at all during the lifetime of state 
$\ket{-}$. As will be seen below, however, such a setup is not very 
convenient for assembling a quantum processor containing many qubits, which 
necessitates the above analysis. It is also clear now that if the same probe 
laser is applied to the qubit for a longer time 
$T_{\text{pr}} \geq \ga_{-+}^{-1}$, it will initialize the state of the 
qubit to its ground state $\ket{G}$.

\section{Entanglement between qubits}
\label{sec:ent}

Universal quantum computation requires the implementation of arbitrary
single-qubit rotations addressed in the previous section and two-qubit
logic gates. Here we will discuss two mechanisms for entangling pairs 
of dimer qubits at well defined locations.

\subsection{Swap gate}

\begin{figure}[t]
\centerline{\includegraphics[width=8.5cm]{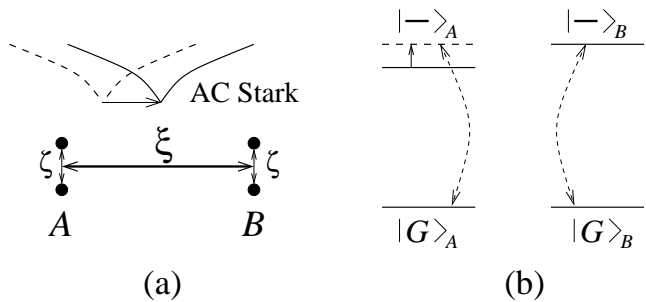}}
\caption{(a) Dimers  $A$ and $B$ are separated by normalized distance
$\xi > \zeta$. An external ac Stark field can switch on and off the RDDI 
between the dimers.
(b) When the qubit transitions of dimers $A$ and $B$ are brought to resonance,
they start swapping a single excitation.
\label{fig:cswp}}
\end{figure}

Consider the RDDI between two dimers $A$ and $B$ separated by normalized 
distance $\xi = q r_{AB}$ satisfying the condition $\zeta < \xi \ll 1$ 
[Fig. \ref{fig:cswp}(a)]. Since during the operation of the quantum computer 
only the qubit states of the dimers are populated, we can simplify our 
treatment of the dimer-dimer interaction by considering only the interaction 
between two two-level systems with the ground and excited states $\ket{G}$ 
and $\ket{-}$. Then, from the above analysis we infer that the real part 
of the RDDI potential between the dimers, responsible for the coherent 
excitation exchange between state $\ket{-}_A$ of dimer $A$ and state 
$\ket{G}_B$ of dimer $B$ and vice versa [Fig. \ref{fig:cswp}(b)], can be 
approximated as 
$\De_{AB}^{(-)} \simeq 3 \Ga_- /(4 \xi^3) = 3 \ga \zeta^2/(20 \xi^3)$.

Let us assume that we have a means to selectively control the frequencies
of transitions $\ket{G} \to \ket{-}$ in both dimers. This can be 
accomplished, for example, by applying a far off-resonant standing-wave 
electric field whose node position coincides with the location of dimer $A$. 
Then at the positions of the two dimers the electric field amplitudes will 
differ and dimers $A$ and $B$ will experience different ac Stark shifts 
[Fig. \ref{fig:cswp}(a)]. If the difference in the qubit transition 
frequencies of the two neighboring dimers exceeds their coupling strength 
$\De_{AB}^{(-)}$, the excitation exchange (\textsc{swap}) between 
them is effectively switched off. To switch the interaction on, one shifts 
the Stark field profile along the $A-B$ axis until qubit transitions of the 
two dimers become resonant. Then, during the time 
$T_{\textsc{swap}} = \pi /(2 \De_{AB}^{(-)})$, the following transformation 
takes place, 
\begin{equation}
\ket{-}_{A(B)}\ket{G}_{B(A)} \to -i \ket{G}_{A(B)}\ket{-}_{B(A)}, 
\end{equation}
while other initial states of the two qubits, $\ket{-}_{A}\ket{-}_{B}$ and 
$\ket{G}_{A}\ket{G}_{B}$, remain unaffected. This is the essence of the
\textsc{swap} gate between two qubits. 

In the same way, one can realize the {\it square-root of} \textsc{swap} 
($\sqrt{\textsc{swap}}$) gate between two qubits. By switching on the 
interaction for time $T_{\sqrt{\textsc{swap}}} = \pi /(4 \De_{AB}^{(-)})$, 
one can fully entangle the two qubits, attaining an equally weighted 
superposition of \textsc{swap} and no-\textsc{swap},
\begin{equation}
\ket{-}_{A(B)}\ket{G}_{B(A)} \to  
\frac{1}{\sqrt{2}} [\ket{-}_{A(B)}\ket{G}_{B(A)} - 
i \ket{G}_{A(B)}\ket{-}_{B(A)}] 
\label{sqrt_swap} .
\end{equation}

Let us estimate the fidelity $F_{\text{swap}}$ of the \textsc{swap} operation.
The main source of error in this scheme is the cooperative spontaneous 
decay of the excited states of the qubits, 
$P_{\text{swap}}^{\text{sp}} \leq 2 \Ga_- T_{\textsc{swap}} = 4 \pi \xi^3/3$. 
With inter-dimer separation $\xi \simeq 0.1 \gg \zeta$, this leads to 
the \textsc{swap} gate fidelity $F_{\text{swap}} = 
1 - P_{\text{swap}}^{\text{sp}} \geq 0.996$. 

We note that the decoherence-free subspace approach \cite{ZB,Lidar} 
advocates the use of four physical qubits (TLAs) for a single logical qubit 
represented by two subradiant states of the four-atom system. The universal 
set of quantum gates relies on the exchange interaction (\textsc{swap}) 
between pairs of atoms located at a logical qubit (for single-qubit rotation) 
or at different logical qubits (for two-qubit gate), which can be turned on 
and off via externally applied electric or magnetic fields. Our dimer qubit
approach allows for more efficient use of the system resources (two atoms
per qubit), along with simpler and more robust manipulation.

\subsection{Fast controlled-phase gate}

\begin{figure}[t]
\centerline{\includegraphics[width=8.5cm]{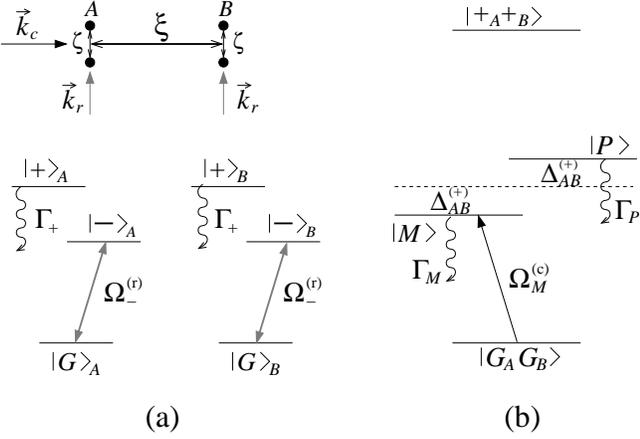}}
\caption{(a) Geometry of the system of two dimers and their internal
level structure. Each qubit can be addressed separately by a laser field 
with $\textbf{k}_r \parallel \textbf{r}_{12}$. Two-qubit 
interaction is mediated by a coupling field with 
$\textbf{k}_c \parallel \textbf{r}_{AB}$.
(b) Eigenstates of the combined system of two dimers.
\label{fig:cphs}}
\end{figure}

We now describe an alternative scheme implementing a fast controlled-phase 
(\textsc{cphase}) logic gate between two qubits [Fig. \ref{fig:cphs}]. 
Suppose that we irradiate the dimers with a laser field acting on 
the auxiliary transition $\ket{G} \to \ket{+}$, and thereby populating the 
state $\ket{+}$. This will induce the RDDI between two closely spaced 
dimers causing an excitation exchange between state $\ket{+}_A$ of dimer 
$A$ and state $\ket{G}_B$ of dimer $B$ and vice versa. Using the above 
analysis, we obtain that the strength of the interaction is 
given by $\De_{AB}^{(+)} \simeq 3 \Ga_+ /(4 \xi^3) = 3 \ga /(2 \xi^3)$, 
which is much larger than $\De_{AB}^{(-)}$, since 
$\Ga_+/\Ga_- \simeq 10/\zeta^2 \gg 1$. Therefore, during a time interval 
that is small compared to $|\De_{AB}^{(-)}|^{-1}$, we can neglect the RDDI 
between the dimers on the qubit transitions $\ket{G}_{A,B} \to \ket{-}_{A,B}$ 
in comparison to that on the auxiliary transitions 
$\ket{G}_{A,B} \to \ket{+}_{A,B}$. To the same accuracy, the eigenstates 
of the combined system of two dimers are given by 
\[
\begin{array}{c}
\ket{G_A G_B} , \;\;\;\;  \ket{+_A +_B} , \\ \\
\ket{M} = \frac{1}{\sqrt{2}} (\ket{+_A G_B} - \ket{G_A +_B}), \\ \\
\ket{P} = \frac{1}{\sqrt{2}} (\ket{+_A G_B} + \ket{G_A +_B}) .
\end{array}
\]
Thus, the singly excited states $\ket{M}$ and $\ket{P}$, having the decay 
rates $\Ga_M \simeq \Ga_+ \xi^2/5$ and $\Ga_P \simeq 2 \Ga_+$, correspond, 
respectively, to the antisymmetric and symmetric combinations of the 
superradiant $\ket{+}$ and ground $\ket{G}$ states of two dimers
[Fig. \ref{fig:cphs}(b)]. 

The geometry of the system is depicted in Fig. \ref{fig:cphs}(a), where
the interatomic axis of each dimer is perpendicular to the inter-dimer axis,
$\textbf{r}_{12}^{A,B} \perp \textbf{r}_{AB}$. Each qubit can be separately 
addressed by the laser field with $\textbf{k}_r \parallel \textbf{r}_{12}$,
as described in Sec.~\ref{sec:sqr}. To perform a two-qubit logic gate, we 
irradiate the system with the coupling field 
$E_c = \mathcal{E}_c e^{i (\textbf{k}_c \textbf{r} -\om_c t)}$ having 
wave vector $\textbf{k}_c \parallel \textbf{r}_{AB}$ and frequency 
$\om_c = \om_{eg} + \De - \De_{AB}^{(+)}$ that is resonant with the 
transition $\ket{G_A G_B} \to \ket{M}$. The Rabi frequencies of this field on 
the transitions $\ket{G_A G_B} \to \ket{M}$ and $\ket{G_A G_B} \to \ket{P}$ are
equal, respectively, to $\Om_M^{(c)} = \Om_c \xi$ and $\Om_P^{(c)} = 2 \Om_c$.
Since $\textbf{k}_c \perp \textbf{r}_{12}^{A,B}$, this field does not
couple to the qubit transitions of the dimers. During the time 
$T_{\textsc{cphase}} = \pi/\Om_M^{(c)}$, the system of two dimers, being 
initially in the state $\ket{G_A G_B}$, will undergo one Rabi cycle on the
transition $\ket{G_A G_B} \to \ket{M}$ and the following transformation will 
take place,
\begin{equation}
\ket{G_A G_B} \to - \ket{G_A G_B}, \label{cphase}
\end{equation} 
while all other initial states, such as $\ket{-}_{A}\ket{-}_{B}$ and 
$\ket{-}_{A(B)}\ket{G}_{B(A)}$, will remain unaffected. This is due to the 
fact that the RDDI between the dimers is present only if their combined 
state is either $\ket{G}_{A}\ket{+}_{B}$ or $\ket{+}_{A}\ket{G}_{B}$. 
Otherwise there is no resonance in the system corresponding to the frequency 
$\om_c$ of the coupling field and, consequently, the field does not interact 
with the system. Transformation (\ref{cphase}) results in a $\pi$ phase 
shift of the state $\ket{G_A G_B}$ which corresponds to the \textsc{cphase} 
logic gate. The comparison of operation times of the \textsc{swap} and 
\textsc{cphase} gates yields  
$T_{\textsc{swap}}/T_{\textsc{cphase}} = 10 \Om_c \xi^4/(3 \ga \zeta^2)$.

We now estimate the fidelity of the \textsc{cphase} gate. The first
source of error is the spontaneous emission from state $\ket{M}$, 
$P_{\textsc{cphase}}^{\text{sp}} \leq \Ga_M T_{\textsc{cphase}}=
2 \pi \ga \xi/5 \Om_c$. Next, an error may occur if the coupling field 
transfers some population from the ground $\ket{G_A G_B}$ to the excited 
$\ket{P}$ state of the system, from where it will decay back to the ground 
state with random phase, $P_{\textsc{cphase}}^{\text{tr}} \leq 
\Ga_P |\Om_P^{(c)}|^2 T_{\textsc{cphase}} /(2 \De_{AB}^{(+)})^2 = 
16 \pi \Om_c \xi^5/(9 \ga)$. The last source of error comes about when
only one of the qubits is in the ground state $\ket{G}$ and, therefore, the 
\textsc{cphase} gate is not executed. However, the application of the 
$\mathcal{E}_c$ field to that qubit may result in a small population transfer 
to the superradiant dimer state $\ket{+}$. The probability of that process, 
$\tilde{P}_{\textsc{cphase}}^{\text{tr}} \leq \Ga_+ |\Om_+^{(c)}|^2 
T_{\textsc{cphase}} /(\De_{AB}^{(+)})^2$, turns out
to be equal to $P_{\textsc{cphase}}^{\text{tr}}$. Minimizing the total
error probability $P_{\textsc{cphase}} = P_{\textsc{cphase}}^{\text{sp}}
+ P_{\textsc{cphase}}^{\text{tr}}$ with respect to $\Om_c$, we have
$P_{\textsc{cphase}}^{\text{min}} \leq 8 \ga/\De_{AB}^{(+)} \simeq 
5.3 \xi^3$ for $\Om_c /\ga = (2 \xi^2)^{-1}$.  With $\xi \simeq 0.1 > \zeta$ 
and $\Om_c/\ga \simeq 50$, we obtain for the \textsc{cphase} gate fidelity
$F_{\textsc{cphase}} = 1- P_{\textsc{cphase}}^{\text{min}} \geq 0.995$,
which is similar to that of the \textsc{swap} gate. However, for
the chosen parameters, the \textsc{cphase} gate is 15 times faster than 
the \textsc{swap} gate.

We note finally that a related scheme implementing the \textsc{cphase} logic 
gate between two closely spaced Raman qubits (Fig. \ref{fig:R_tr}) have been 
proposed in \cite{BCJD}. In that scheme, one applies to the pair of atoms
$A$ and $B$, trapped in an optical lattice, a ``catalysis'' field 
$\mathcal{E}_C$ having frequency that is near-resonant with the atomic 
transition $\ket{g_2} \to \ket{e}$. The detuning of that field $\de_e^{(C)}$ 
is smaller than the splitting of the ground states $\ket{g_1}$ and $\ket{g_2}$ 
but larger than the RDDI strength $\De_{AB}^{(R)} \simeq 3 \ga_e/(4 \xi^3)$ 
between the atoms on the transitions $\ket{g_2}_{A,B} \to \ket{e}_{A,B}$. 
Therefore the RDDI is induced only if both atoms are in state $\ket{g_2}$. 
During the interaction with the catalysis field, different initial states of 
the system, $\ket{g_1}_{A} \ket{g_1}_{B}$, $\ket{g_1}_{A(B)} \ket{g_2}_{B(A)}$,
and $\ket{g_2}_{A} \ket{g_2}_{B}$, experience the corresponding ac Stark 
shifts $S_{g_1 g_1} = 0$, $S_{g_1 g_2} = 2|\Om_C|^2/\de_e^{(C)}$ and
$S_{g_2 g_2} = 2|\Om_C|^2/[\de_e^{(C)}-\De_{AB}^{(R)}]\simeq 
S_{g_1 g_2} [1+\De_{AB}^{(R)}/\de_e^{(C)}]$. Thus, to perform 
the \textsc{cphase} gate, one applies the catalysis field for time 
$T_{\textsc{cphase}}^{(R)} = \pi / (S_{g_2 g_2} - S_{g_1 g_2}) = 
\pi \de_e^{(C)}/[\De_{AB}^{(R)} S_{g_1 g_2}]$. However, the single-atom 
phase-sifts $S_{g_1 g_2} T_{\textsc{cphase}}^{(R)}$, accumulated during the 
gate operation, should be removed through appropriate pulses acting on 
individual atoms before and after they are made to interact. The probability 
of error due to the spontaneous emission from the excited states 
$\ket{e}_{A,B}$ is given by $P_{\textsc{cphase}}^{(R)} \simeq 
2 \ga_e S_{g_1 g_2} T_{\textsc{cphase}}^{(R)} /\de_e^{(C)} = 8 \pi \xi^3/3$.
With $\xi \simeq 0.1$, which here has a meaning of the Lamb-Dicke parameter, 
the error probability $P_{\textsc{cphase}}^{(R)} \simeq 8 \times 10^{-3}$ 
is slightly larger than in our scheme. More dramatically, however,
for similar field strength $\Om_C/ \ga_e \simeq 50$ and 
$\de_e^{(C)} \simeq 5 \De_{AB}^{(R)} \gg \Om_C$, we find that our 
implementation of the \textsc{cphase} gate, in addition to being simpler, 
is $T_{\textsc{cphase}}^{(R)}/T_{\textsc{cphase}} \simeq 20$ times faster.

\section{Implementation of quantum processor}
\label{sec:impl}

Having established all the basic physical principles of operation of the 
proposed quantum processor, we now describe its possible realization. The 
processor is composed of a solid-state host doped with active atoms. These 
atoms should have a non-degenerate ground state, since otherwise the coupling 
between atoms via vacuum modes of the continuum can mix various degeneracy 
states \cite{GKABR}, which would invalidate the simple two-level atomic 
model we have explored in this paper. Among the possible dopants, pairs of 
semiconductor quantum dots, often referred to as artificial atoms 
\cite{optQDs}, with controllable separations of few nanometers \cite{BHHFKWSF},
appear to be the best choice for our scheme, due to their large dipole moments 
and tailorable optical properties. 

\begin{figure}[t]
\centerline{\includegraphics[width=8.5cm]{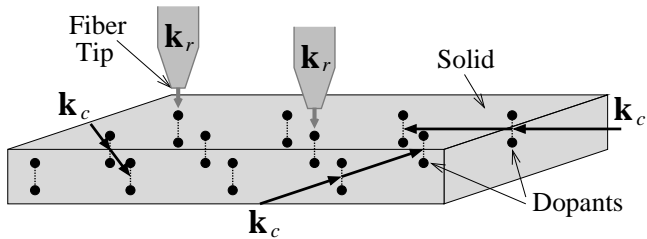}}
\caption{Schematic drawing of the proposed quantum processor with RDDI 
dopants in solid state host.
\label{fig:q_pr}}
\end{figure}

In Fig. \ref{fig:q_pr} we show the scheme of the proposed quantum processor
with RDDI dopants in solid state host. There, the qubits are represented by 
the ground and subradiant states of the dimers formed by pairs of closely 
spaced atoms. Individual qubits are addressed by laser fields with frequency 
$\om_r$ and wave vector parallel to the interatomic axis, 
$\textbf{k}_r \parallel \textbf{r}_{12}$, using the near-field technique. 
The polarization of the field can be chosen such that it acts only on the 
atomic transition from the nondegenerate ground state to one of the sublevels 
of the excited state. This further justifies the validity of the two-level 
description of the atoms.

Throughout this paper we assumed that the relaxation of the excited atomic  
state $\ket{e}$ is conditioned merely by the radiative decay. This assumption 
has resulted in a strong suppression of decoherence on the qubit transition 
due to the subradiant nature of the antisymmetric state of the dimer. For this 
to be valid, during the operation of the quantum processor all the competing 
nonradiative decay processes must be strongly suppressed. This can be 
accomplished by working below the liquid helium temperatures, at which the 
density of crystal phonons is negligible \cite{phonons}, and/or using fast 
ac Stark modulation of the vibrationally relaxing levels \cite{zeno}. Another 
important issue that has to be briefly addressed here is the influence of the 
inhomogeneous broadening of the atomic resonances. Consider two near-RDDI 
atoms having slightly different resonant frequencies, 
$\om_{eg}^{(2)} - \om_{eg}^{(1)} = \de \om_{eg} \ll \om_{eg}^{(1,2)}$, 
due to, e.g., size inhomogeneity of the quantum dots and/or local defects 
of the host material. One can show that this frequency mismatch results in 
an increase of the decay rate $\Ga_-$ of the qubit excited state $\ket{-}$, 
given by $\ga \, \de\om_{eg}^2/(8 \De^2)$. 
If we require that this additional relaxation rate does not exceed $\Ga_-$ 
for two resonant atoms, we obtain the following condition on the width of 
the inhomogeneous broadening $\de\om_{eg} \leq \ga/\zeta^2$, which, for  
interatomic separations $\zeta \sim 2-5$ nm and optical or near infrared 
transitions, is by three orders of magnitude larger than the width of  
homogeneous broadening $\ga$ of the atomic transition $\ket{g} \to \ket{e}$.

It is known that using a suitable sequence of one-qubit rotations and 
two-qubit entanglement one can obtain any desired unitary transformation of 
the system \cite{QCQI}. With the arrangement of dimers shown 
in Fig. \ref{fig:q_pr}, our scheme is capable of implementing two different 
two-qubit logic gates. The \textsc{cphase} gate between two qubits $A$ and 
$B$ is executed by a coupling field whose wave vector points in the direction 
of the interqubit axis, $\textbf{k}_c \parallel \textbf{r}_{AB}$. Since, in 
general, for any pair of qubits the vector $\textbf{r}_{AB}$ is different, 
the frequency $\om_c$ of the coupling laser is also different, which 
facilitates selective entanglement of a chosen pair of qubits $A$ and $B$.
The \textsc{swap} gate between neighboring qubits is always present. It can
be used to convey the information in the quantum processor, step-by-step 
from one qubit to another, between the qubits that are separated by large 
distances, over which the direct RDDI between them vanishes. To neutralize 
the \textsc{swap}, one can flip the qubits at time intervals that are short 
compared to $[\De_{AB}^{(-)}]^{-1}$, which is equivalent to the spin echo 
technique used in NMR \cite{NMR}. Otherwise the $\sqrt{\textsc{swap}}$ gate 
between two qubits $A$ and $B$ can be switched on and off via external ac 
Stark fields. Finally, the readout is performed by shining at the qubit a 
probe laser with the frequency $\om_p$ and detecting the fluorescence if 
the qubit state is $\ket{G}$. The same probe laser, if shined at the qubit 
for a time longer than $\ga_{-+}^{-1}$, will initialize the state of the 
qubit to its ground state $\ket{G}$. 
 
To conclude, we have proposed a realization of a quantum processor using
near-resonant dipole-dipole interacting dopants in a solid state host. 
We have shown that the ground and long-lived subradiant states of the 
effective dimers, formed by pairs of closely spaced two-level systems, can 
serve as reliable qubit states. A robust measurement scheme of the qubit, 
based on the electron shelving technique, has also been discussed. The 
two-qubit entanglement can be realized either by coherent excitation 
exchange between the dimers, or by coupling the qubits via external 
laser fields. 

We have also compared our scheme with other schemes proposed in the 
literature and have shown that the present scheme offers reliable 
single- and two-qubit quantum gates. Another noteworthy advantage is 
that our system is capable of realizing practically important quantum 
computation that requires large number of qubits, which is known to be 
hardly achievable in ion trap \cite{iontr}, cavity QED \cite{PGCZ,MBBKNSK}, 
or NMR \cite{NMR} based schemes. Although our proposal for solid-state 
quantum processor relies on significant experimental advances in 
nanofabrication technology, there are no principle limitations on 
the scalability of this scheme.

\end{document}